\documentclass[%
 aip,
 sd,%
 amsmath,amssymb,
 reprint,%
]{revtex4-1}

\usepackage{graphicx}
\usepackage{dcolumn}
\usepackage{bm}
\usepackage{subfigure}
\usepackage{ulem,amsfonts}

\begin{document}

\preprint{AIP/123-QED}

\title[]{Mode Control in a Multimode Fiber Through Acquiring its Transmission Matrix  from a Reference-less Optical System}

\author{Moussa N'Gom}
 \email[corresponding author] {: mngom@umich.edu}
 \affiliation{University of Michigan \\Department of Electrical Engineering \& Computer Science \\ 1301 Beal Ave, Ann Arbor, MI 48109}
\author{Theodore B. Norris}%
\affiliation{University of Michigan \\Department of Electrical Engineering \& Computer Science  \\ 1301 Beal Ave, Ann Arbor, MI 48109
}%

\author{Eric Michielssen}
\affiliation{%
University of Michigan \\Department of Electrical Engineering \& Computer Science  \\ 1301 Beal Ave, Ann Arbor, MI 48109
}%

\author{Raj Rao Nadakuditi}
\affiliation{%
University of Michigan \\Department of Electrical Engineering \& Computer Science  \\ 1301 Beal Ave, Ann Arbor, MI 48109
}%

\date{\today}

\begin{abstract}
A simple imaging system together with complex semidefinite programming is used to generate the transmission matrix of a multimode fiber.  Once the transmission matrix is acquired, we can modulate the phase of the input signal to induce strong mode interference at the fiber output.
The  optical design does not contain a reference arm and no interferometric measurements are required. We use a phase-only spatial light modulator to shape the profile of the propagating modes and the output intensity at an individual pixel is monitored. The semidefinite program uses a convex optimization algorithm to generate the transmission matrix of the optical system using intensity only measurements.
This simple yet powerful method can be used to compensate for modal dispersion in multimode fiber communication systems. It also yields great promises for  the next generation biomedical imaging, quantum communication, and cryptography.
\end{abstract}

\pacs{Valid PACS appear here}
\maketitle
A monochromatic light beam launched into a multimode fiber (MMF) typically excites many  modes in the waveguide. The set of modes excited depends on the coupling conditions at the input of the fiber. Each guided mode travels down the waveguide with a different group velocity. As a result, the information carried by each mode arrives at the end of the fiber at a different time. This phenomenon is known as modal dispersion (MD). In a multimode fiber optical communication system, the dominant factor limiting the achievable  bandwidth is interference between channels caused by MD \cite{agrawal}.  Thus precise modal excitation and control is of great interest for improving the transmission capacity of MMF. 
\\
Mode control has been previously demonstrated using binary phase masks to characterize and suppress MD in MMF \cite{carpenter}. Recently, a variety of new methods using digital phase conjugation \cite{papa, farahi} and wavefront shaping techniques \cite{antonio} (and references therein) have shown precise mode control in MMF endoscopes.
Due to unpredictable mode coupling arising from fabrication defects, bends, and temperature fluctuations, MMF endoscopes require rigorous calibration procedures to enable imaging.  
Therefore, implementing a simple and accurate method to control light propagation inside the fiber is desired. 
\\
When the phase delay among multiple propagating modes varies over the entire range of 2$\pi$ radians, highly structured interference effects or speckle patterns are detected at the output of the fiber \cite{goodman}, analogous to those observed in turbid media. 
For this reason, many approaches to wavefront shaping through highly scattering media \cite{ivo} can be applied to light propagation in MMF. These include the determination of the  transmission matrix (TM) in linear scattering systems \cite{popoff, gigan}. 
\\
The TM of a MMF connects the modes of the input signal coupled into the fiber to the outgoing modes collected at the output of the fiber. Since the number of modes in a MMF is well defined and the numerical aperture (NA) is limited, the fiber TM  can be completely measured. Complete knowledge of the TM enables unprecedented control of propagation through MMF.  
\\
Knowledge of the TM have been used to modulate the signal at the fiber input to generate  sharp foci at the fiber output \cite{silvio, cizmar, papa}, and to transmit images \cite{silvio1, choi, joel} through the fiber. The TM method has also been used to design robust and light efficient single MMF endoscopes \cite{yugu, antonio}.
\\ 
Measuring the TM has also been proposed as an approach to controlling modal dispersion \cite{milione} by enabling selective launching of the principal modes of the fiber \cite{cao}. The principal modes in a MMF form an orthogonal basis at both the input and output of the waveguide and do not suffer from MD \cite{fan}.
The TM has also been used to control the propagation of two photon or quantum light through a MMF \cite{hugo}.
\\
Most techniques to date use  variations of the interferometric method to generate the transmission matrix of the MMF; however, in a telecommunication environment where MMF play a major role for data transport, these techniques can be very difficult to implement. 
Therefore, methods to determine the TM without the need for a reference beam would be tremendously useful, and the development of algorithms for signal recovery from magnitude measurements in a realistic telecommunication system is desirable. 
Simulations results, using parameters of commercially available fibers and SLMs, have shown \cite{elad, shen, rahul, rahul09}  that modal dispersion can be eliminated in a MMF optical communication system using adaptive optics and convex optimization, but they do not infer the TM of the optical system.

\begin{figure}[h!]
\begin{center}
\subfigure[]{\label{MMF_speck} \includegraphics[width=3.0in]{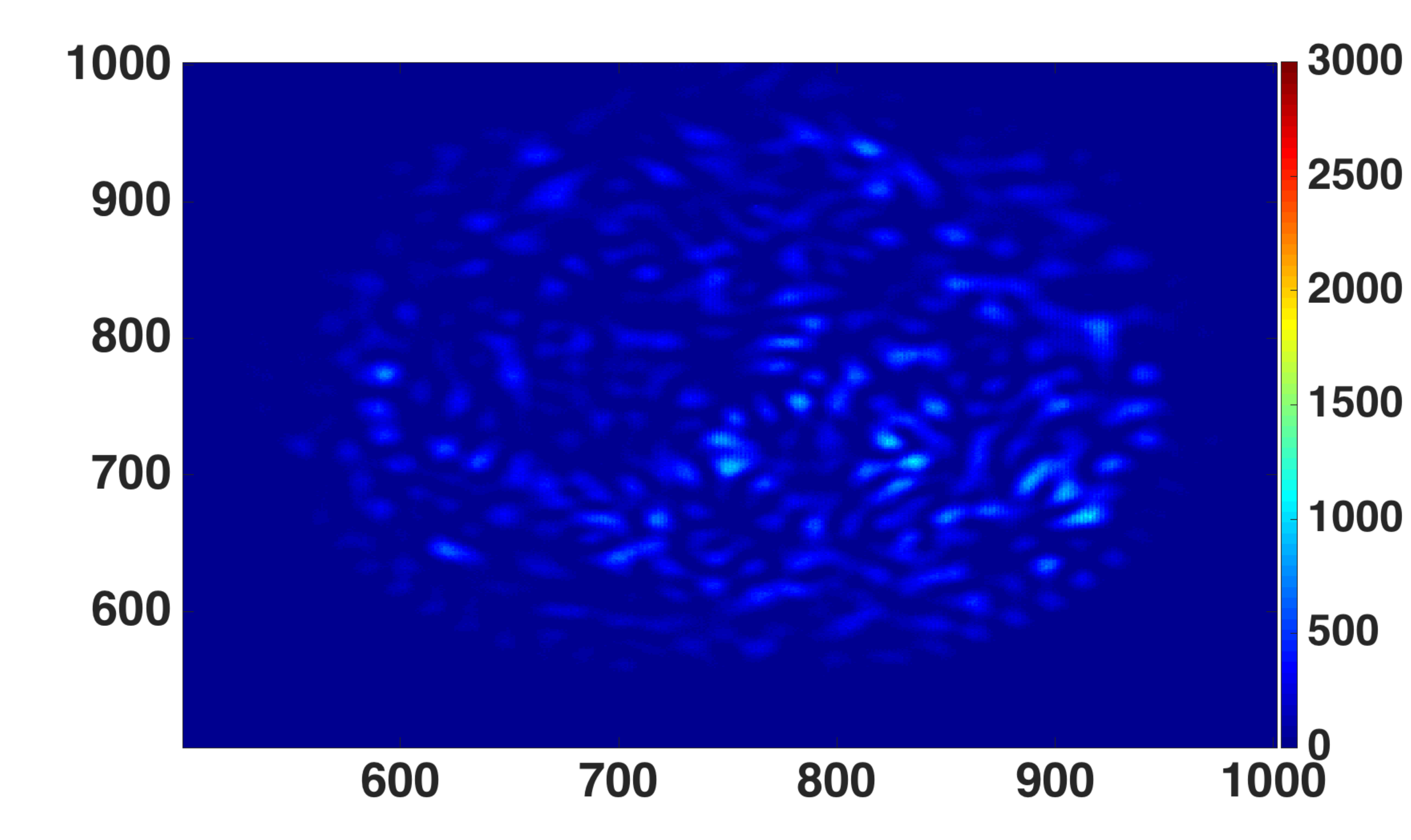}}

\subfigure[]{\label{MMF_focus}			 
			\includegraphics[width=3.0in]{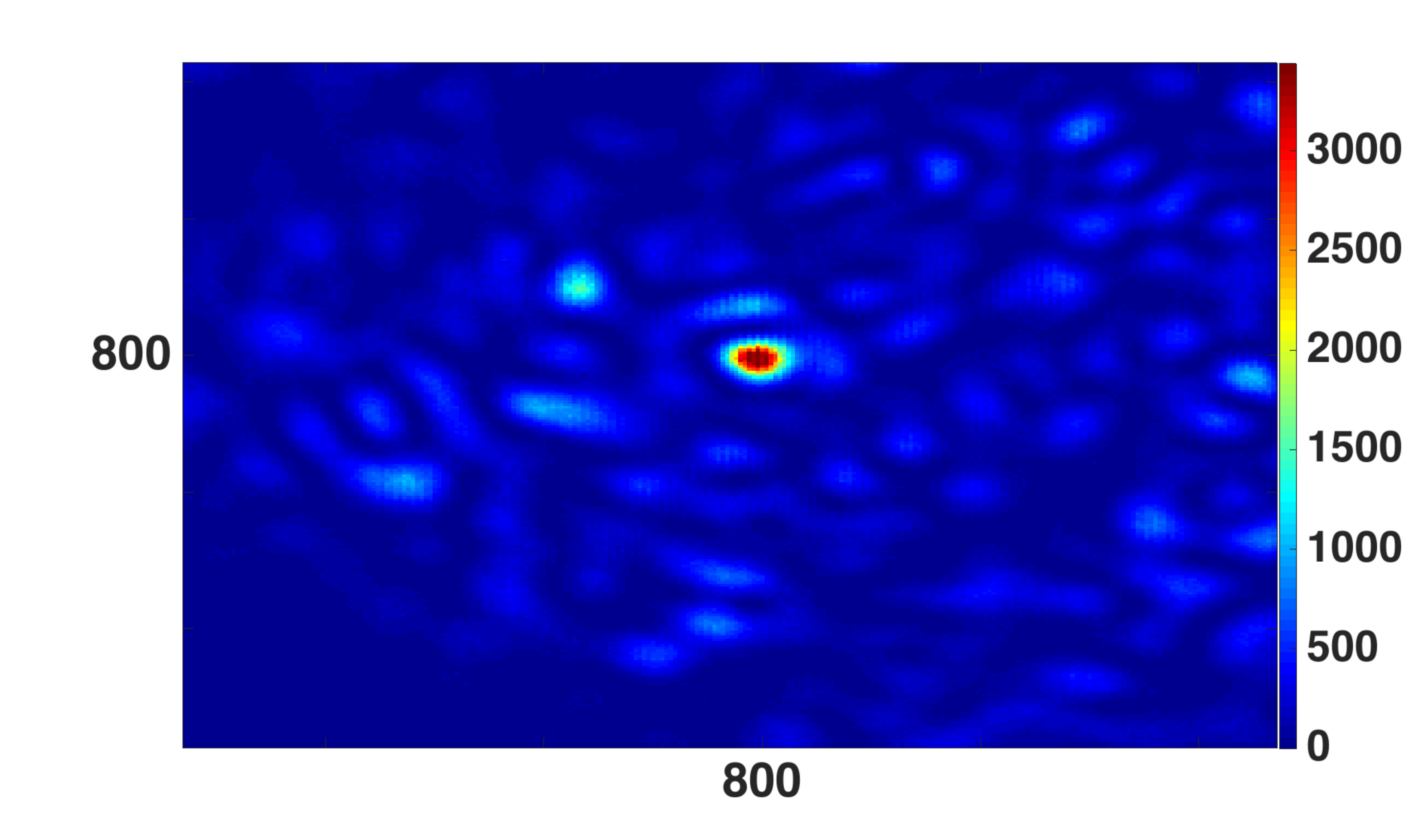}}
\caption{(a) Modal dispersion inside a MMF results in a speckle pattern generated from random wavefront and mode coupling (b) Strong interference of controlled mode after the optimal wavefront is computed from the SDP algorithm.} 
\end{center}
\end{figure}

\begin{figure}[h!]
\begin{center}
\subfigure[]{\label{top} \includegraphics[width=3.0in]{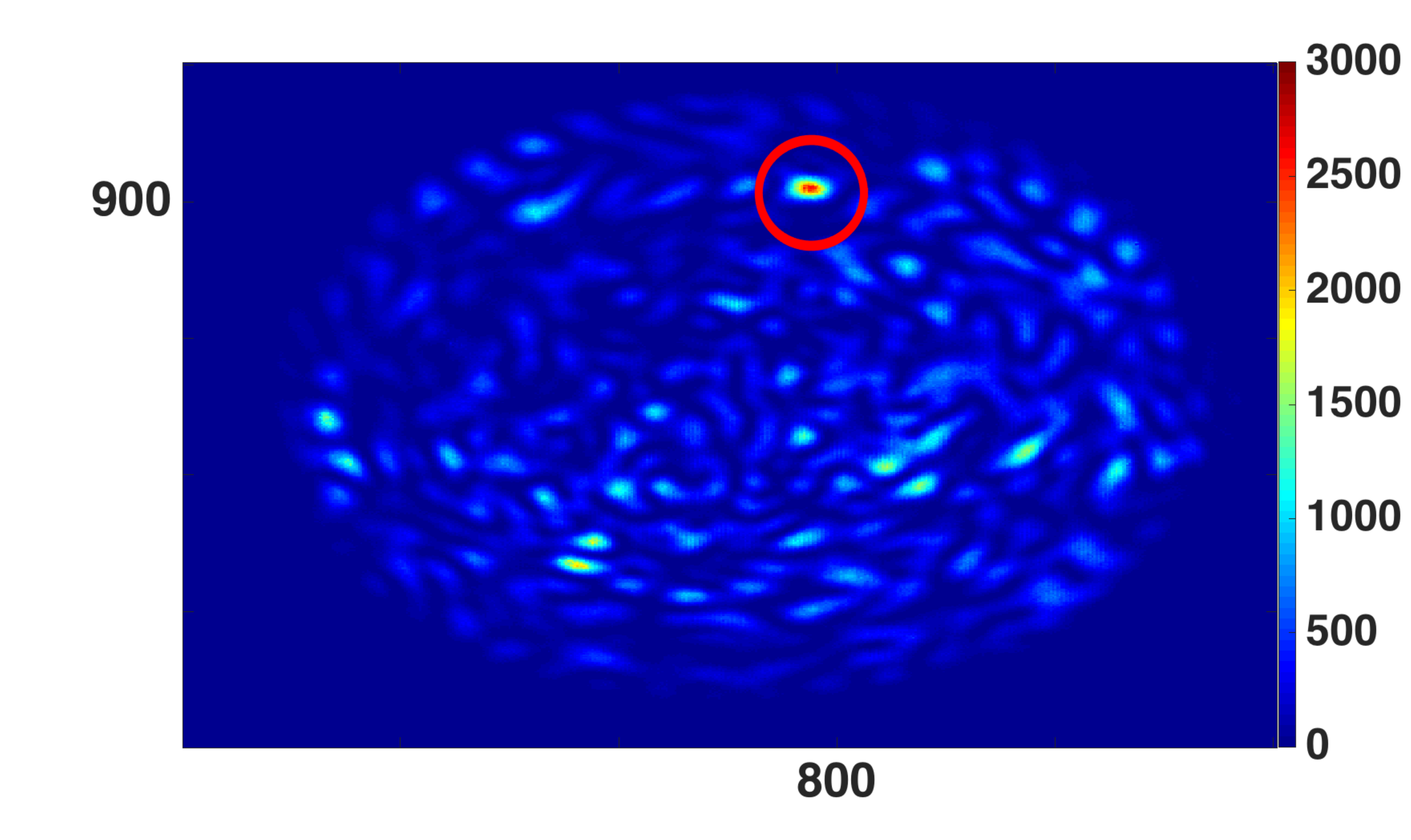}}

\subfigure[]{\label{bottom}			 
			\includegraphics[width=3.0in]{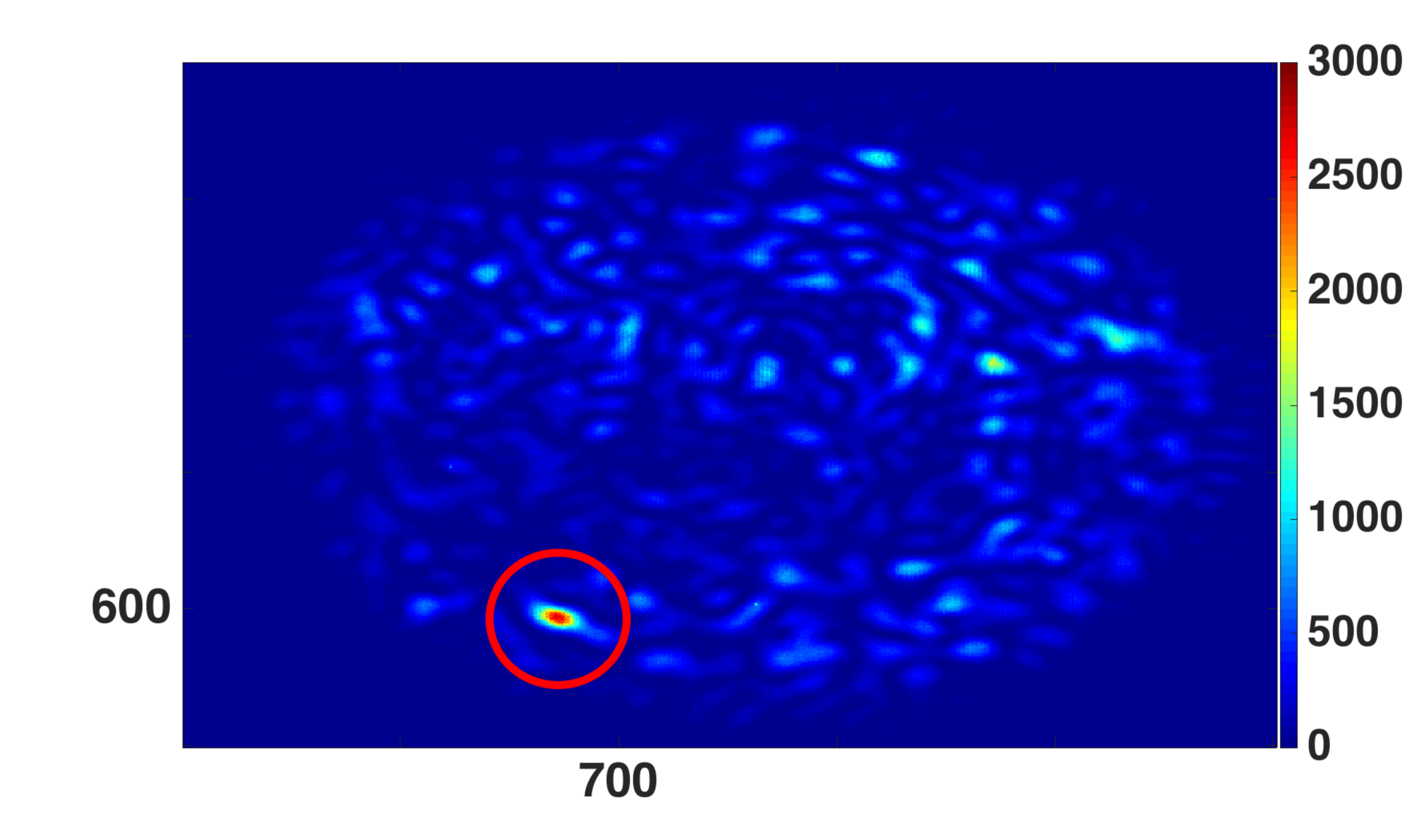}}

\caption{Strong mode interference is induced at different  position at the fiber output. Here we show intense foci at the (a) top (800, 900) pixel location (b) bottom (675, 600) pixel location} 
\end{center}
\end{figure}

In this paper, we  determine the TM of a MMF using a simple optical setup and employing  a semidefinite programming (SDP) algorithm which uses  convex optimization. 
This paper builds on our previous work in \cite{ngom}. The method therein only computes the relevant portion of the transmission matrix where for example a strong focus is desired, rather than generating the `whole' TM. The full TM can also be computed but for efficiency and speed, only the relevant portions are computed.
\\We utilize  a spatial light modulator (SLM) to control the phase of the input signal. The SLM pixels are grouped into a square array of superpixels, where the number $M$ of superpixels is chosen to be  approximately equal to the number of the MMF modes. The phase is constant over each superpixel. The goal is to find the optimum set of phases for each superpixel that produces the desired output, which can be obtained by knowledge of the system's TM.
We adapt the algorithm developed by Waldspurger \textit{et al} \cite{irene} to retrieve the phase of elements of the TM from measurements of the light intensity at the output face of the MMF. 
Hence knowledge of the TM enables the SLM to excite specific  modes while suppressing others. This method will minimize modal dispersion in the system and improve the signal to interference ratio in the MMF channel \cite{elad, rahul09}.
\begin{figure}[h!]
\begin{center}
\includegraphics[width=3.50 in]{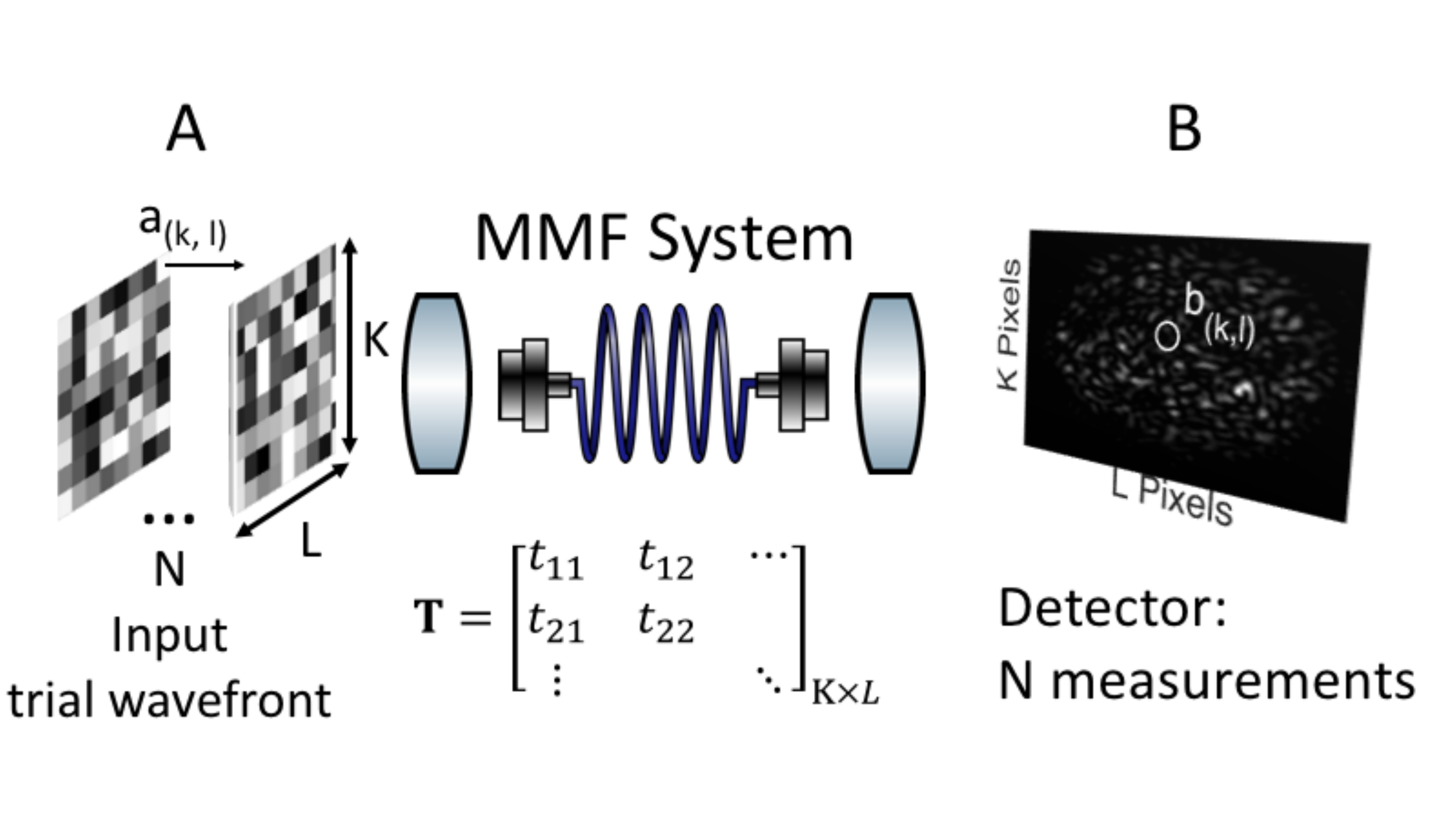}
\caption{Algorithm conceptual setup: the optical system including the MMF is sandwiched between two rectangular apertures ($A$ and $B$) . The corresponding  optical setup is shown in Fig. \ref{opt_des}.}
\label{concept}
\end{center}
\end{figure}
\\
Fig \ref{concept} illustrates the mode control scheme.
The computed transmission matrix $T$ links the input fields $a_{(k, l)}$ at each superpixel to the output fields $b_{(k, l)}$ at the detector.   $T$ incorporates  the propagation from the SLM to the detector including the fiber and the imaging optics. 
Apertures $A$ and $B$, as displayed in Fig \ref{concept}, are assumed to be rectangular and are comprised of  $M = K \times L$ elements.  Light in the input and output apertures are characterized by $M$  complex-valued electric field samples $a_{(k, l)}$ and $b_{(k, l)}$ respectively, with $1 \leq k \leq K, 1 \leq l \leq L$. Our experimental setup only allows for measurements of the magnitude of each $b_{(k, l)}$. 
\\We will next proceed to give an overview of the algorithm, which we have described in detail elsewhere \cite{ngom}. The method therein only computes the relevant portion of the transmission matrix where for example a strong focus is desired rather than generating the `whole' TM.
\begin{figure}[h!]
\begin{center}
\includegraphics[width=3.5in]{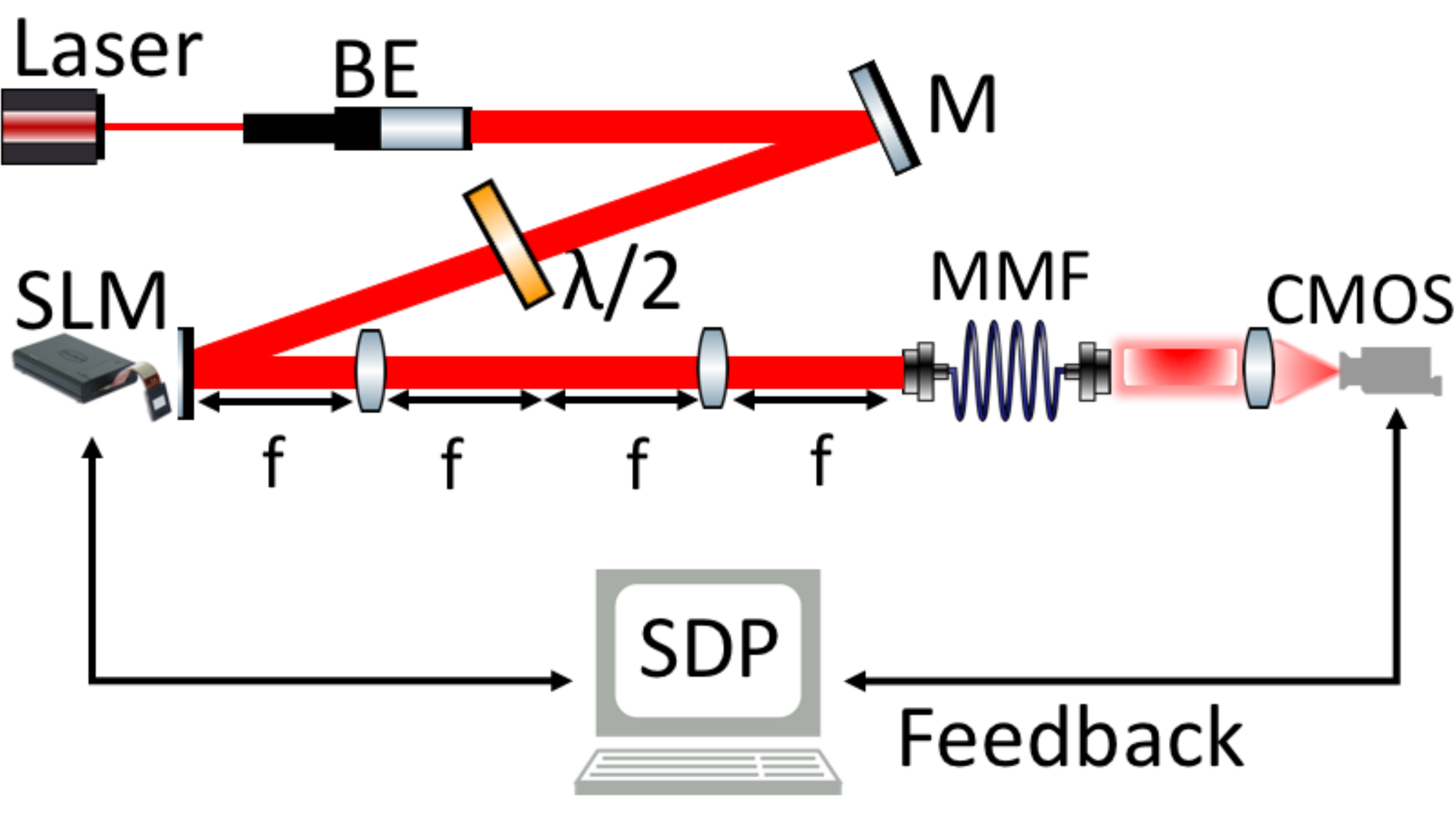}
\caption{A 633 nm CW laser beam is expanded then reflected from the SLM, which is $4f$ imaged and coupled onto the MMF. The output signal from the fiber is collected and imaged onto the CMOS camera.}\label{opt_des}
\end{center}
\end{figure}
\\The fields in $A$ and $B$  are connected in amplitude and phase by the TM $(T)$ of the optical system, 
\begin{equation}\label{Efields}
b = T \cdot a
\end{equation}
 where vector $a = [ \cdots,  a_{(k, l)} ,  \cdots]^{T}$ and $ b = [ \cdots,  b_{(k, l)} ,  \cdots ]^{T}$. The TM in Eq. \ref{Efields} is usually generated by determining the amplitude and phase of the output fields by an interferometric measurement (see reference \cite{gigan} for a comprehensive review). 
In an actual fiber telecommunication network however, establishing an interferometric system is highly nontrivial, motivating the need to determine the TM of the system without  requiring a reference arm.
 \\
 Intensity measurements only allow for the observation of 
 \begin{equation}\label{phase retrieval model}
 |b|^2 = |T \cdot a|^2, 
 \end{equation}
 so the phase of $T$ is undetermined, necessitating an approach to phase retrieval.
 Solving for $T$ is a phase retrieval  problem, which can be divided into sub-problems of solving each row of the TM $(t_{(k, l)})$. Each row corresponds to a pixel $(k, l)$ on the detector. In a telecommunication system each pixel in the detector represents a channel, and each row of the TM is assigned to an individual receiver in the system.
 \\
 The task is to compute an incident field that `forces' the modes of the fiber to constructively interfere  at a predefined location at the output. Mathematically, the goal is to determine the input field vector $a$ that maximizes $|b_{(k, l)}|$ for a predefined $(k, l)$ subject to $||a||_2^2 = 1$.  Only the input phases  are controlled, i.e. $|a_{(k, l)}|^2 = 1/M$ for all $(k, l)$. 
 \\
 To solve for the row of $T$ for a predefined $(k, l)$, we generate $N$ random input fields or training wavefronts $a^{(j)}$ with $j = 1,\cdots, N$. All input fields have uniform magnitude across the input aperture, and the phases  of the input modes are varied randomly between $[0, 2\pi]$.
 \\Recovering T, given $b$ and $a$, as in  Eq \ref{phase retrieval model} is a non-convex problem. By recasting this problem as a SDP, which we accomplish by adopting PhaseCut \cite{irene} as in \cite{ngom}, we  transform the transmission matrix recovery problem in a convex optimization problem. 
\\
It has been shown that, if the number of measurements  $N$ is of order $O(M \log M)$,   then PhaseCut will recover the phase vector  with  high accuracy in a noise-free setting when the training wavefronts are generated at random \cite{irene}, as we have. 
\\
The phase-only   incident field vector $a$ that maximizes $ |b_{(k, l)}| = |t_{(k, l)} \cdot a|$ at a predefined channel $(k, l)$ at the output of the fiber corresponds to:
\begin{equation}\label{focusing_phase}
a_{{\sf opt}} = \arg \max_{ |a_{(k, l)}|^2 =1/M,}  |t_{(k, l)} \cdot a|.
\end{equation}
Eq. (\ref{focusing_phase}) has a closed-form solution that is given by
\begin{equation}\label{eq:aopt_2}
a_{{\sf opt}}= \dfrac{1}{\sqrt{M}} \exp(i \arg[  t_{(k, l)}]),
\end{equation}
where
$$ \exp(i \arg[t_{(k, l)}]) = \begin{bmatrix} \exp\left(i \arg[ t^{(1)}_{(k, l)}]\right) \\ \hdots \\ \exp\left(i \arg[ t^{(M)}_{(k, l)]}]\right) \end{bmatrix}.$$
\\
Fig. \ref{opt_des} shows the simple optical design we use to measure the TM of the MMF system. 
In the experiment, we use a 2 meter long step index fiber  (Thorlabs M50L02S-A), with a numerical aperture (NA) = 0.22. The  core diameter of  the fiber is $ d = 50 \mu m$. The number of supported modes is given by $m \approx \frac{V^2}{2} $ in each polarization direction, where $V$ is the normalized optical frequency, also known as the V-number of the fiber, $V = \frac{2\pi}{\lambda_o} (NA) \frac{d}{2}$.
The light source is a single longitudinal mode diode laser (CrystalLaser) producing  50 $mW$ at $\lambda \approx 633$ $nm$.
The output beam of the laser is expanded by a beam collimator and reflected onto the SLM (Holoeye PLUTO), which is a  liquid crystal on silicon (LCOS) micro-display with full HD resolution (1920 x 1080 pixel) and 8 $\mu m$ pixel pitch. 
A $\lambda/2$ waveplate and a polarizer are used to control the polarization state of the incident beam to achieve phase only modulation.  
The surface of the SLM is $4f$ imaged onto the entrance face of the fiber. The SLM display is subdivided into $M$  superpixels, i.e. each superpixel contains an equal number of SLM pixels that are modulated together. 
The output signal from the MMF is imaged onto a CMOS camera (PhotonFocus camera series MV1-D2048 with 2048 X 2048 resolution and pixel size 8 $\mu m$). 
 \begin{figure}[h!]
\begin{center}
			\includegraphics [width = 3.5 in]{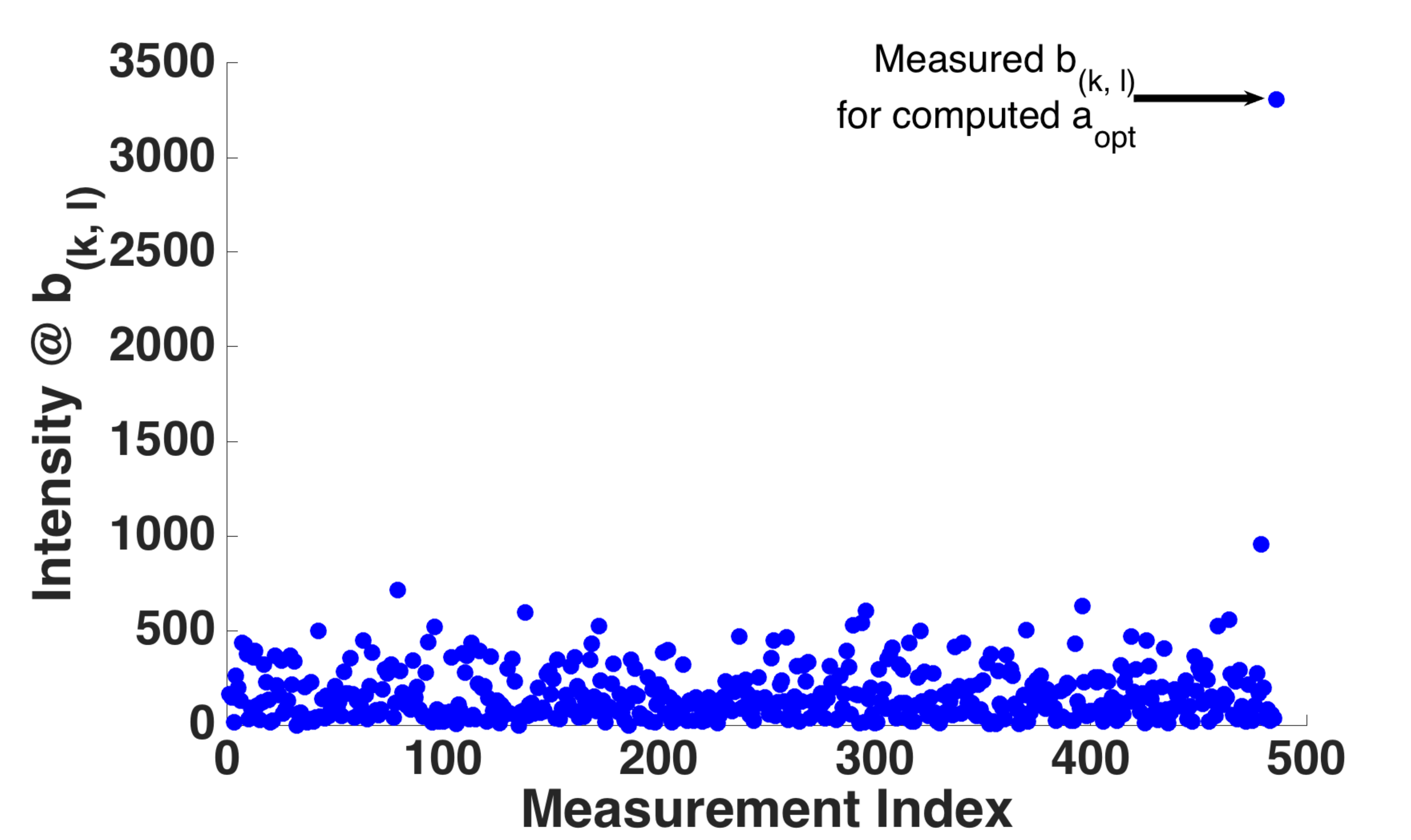}
\caption{Intensity measurements $b$ at pixel $(k, l)$ for every random input wavefront. The computed optimal wavefront ($a_{opt}$) generates the strong interference at the targeted output on the detector} \label{intensity}
\end{center}
\end{figure}
\\The experimental results presented here are obtained by subdividing the SLM into $M = 110$ superpixels. A Sequence of $N = 479 \approx M \log M$ random training wavefronts are injected into the MMF. A  $\lambda /2$ waveplate is used to adjust the polarization of the incident light to match that of the desired fiber modes. The fiber is coiled tightly to induce randomized mode coupling.
 For each incident wavefront $a_{(k, l)}$ the superpixels' phases are randomly set between $[0, 2\pi]$; their corresponding transmitted intensities $b$ are measured at pixel $(k, l)$. Fig. \ref{intensity} shows the intensity output response for the training wavefronts coupled into the MMF input. 
Each training wavefront generates its own unique intensity or speckle pattern.  Fig. \ref{MMF_speck} shows a typical speckle pattern that emerges from a MMF.
 The intensity measurements are collected, transformed, and fed into the SDP program, which then computes the relevant row $t_{(k, l)}$ of the TM. 
 \\
 The next objective is to couple into the fiber the optimal wavefront $a_{{\sf opt}}$  from Eq. \ref{eq:aopt_2} that forces the  propagating modes to interfere at a predetermined position. 
Fig. \ref{intensity} shows that the computed optimal wavefront generates  the highest intensity measurement at the targeted position on the detector. Fig. \ref{MMF_focus} displays the strong interference of the controlled modes near the center of fiber output. The size of the focus spot is determined by the N.A. of the optical system. The enhancement factor at the focus is defined as $\eta$: 
\begin{equation}
\eta = \frac{|b_{(k, l)}|^2}{|\overline{b}_{(k, l)}|^2}
\end{equation}
where $b_{(k, l)}$ is the intensity at $(k, l)$ when $a_{{\sf opt}}$ is coupled into the MMF  and $\overline{b}_{(k, l)}$ is the average intensity at pixel $(k, l)$ over the $N$ training realizations. The strong interference is displayed in Fig. \ref{MMF_focus}, this  yields an enhancement factor of $\eta = 35$.
\\The data clearly demonstrates that the modes coupled in the fiber can be controlled and redirected or refocused into a useful transmitted mode. 
In principle this method allows the controlled modes to interfered anywhere at the fiber output or to any channel in a telecommunication system. For example 
Fig. \ref{top} and \ref{bottom} show interference spots at other locations obtained from the same TM measurements but with different optimal wavefront $a_{{\sf opt}}$. In other words, a single TM enables focusing the signal at any desired output location. 
 \\
 SDP together with faster SLM systems (MEMS and/or micromirrors) could constitute an adaptive optics approach to compensate for modal dispersion in telecommunication.
 This design would provide dynamic modal compensation when the state of the modes inside a MMF evolves as the fiber changes in response to slow time-varying perturbations, e.g., temperature and vibration. 

%
%
In summary, we have demonstrated a non-holographic method to compute the TM of a multimode optical fiber system. 
Obtaining the complete transmission matrix of a fiber system without the use of a reference arm will be an important milestone in the burgeoning field of wavefront shaping. The simplicity of this method has the potential of enhancing the design of single multimode fiber endoscopes, and to improve \textit{in-vivo} characterization in biomedical imaging. 
A great impact can also be made in fiber telecommunication systems with the promise to further increase the bandwidth in MMF by compensating for modal dispersion. 
 Once the TM of the MMF is acquired using this dynamic adaptive optics method, the principal modes of the fiber can be inferred and launched on demand.

\section*{Acknowledgements}
This work was supported by a DARPA Young Faculty Award D14AP00086. The authors also thank the creators and the maintainers of the website \texttt{http://wavefrontshaping.net/} -- the many tips and suggestions and software posted there allowed for this project to come together sooner than it would have otherwise.


\end{document}